%% file: __authordraft.tex
\documentclass[sigconf]{acmart}
\AtBeginDocument{%
  }

\copyrightyear{2026}
\acmYear{2026}
\setcopyright{cc}
\setcctype{by}
\acmConference[UIST '26]{The 39th Annual ACM Symposium on User Interface Software and Technology}{November 02--05, 2026}{Detroit, MI, USA}
\acmBooktitle{The 39th Annual ACM Symposium on User Interface Software and Technology (UIST '26), November 02--05, 2026, Detroit, MI, USA}
\acmDOI{10.1145/3830398.3830645}
\acmISBN{979-8-4007-2856-3/2026/11}

\usepackage{xspace}
\usepackage[capitalise]{cleveref}
\usepackage{tabularx}
\usepackage{booktabs}

\newcommand{\ie}{i.e.\xspace}
\newcommand{\eg}{e.g.\xspace}
\newcommand{\name}{MuTable\xspace}

\usepackage{enumitem}

\usepackage{booktabs}
\usepackage{tabularx}
\usepackage{array}

\definecolor{ModifierFill}{RGB}{255, 247, 230} 
\definecolor{ModifierStroke}{RGB}{214, 212, 207}
\definecolor{ModifierText}{RGB}{46, 44, 39}

\usepackage{acmart-taps}

\usepackage{tikz}

\aptLtoXcmd{%
  \long\def\modifier#1{%
    \xbox{aptbox}{%
      \XMLaddatt{style}{%
        background-color: \#fff7e6;
        color: \#2e2c27;
        display: inline-block;
        padding: 0.05em 0.22em;
        border: 1px solid \#d6d4cf;
        border-radius: 0.25em;
        font-size: 0.85em;
        line-height: 1.15;
        vertical-align: 0.05em;
        box-sizing: border-box;
        white-space: nowrap;
      }%
      #1%
    }%
  }%
}{
\DeclareRobustCommand{\modifier}[1]{\hspace{2pt}{\small{\tikz[overlay]\node[fill=ModifierFill, draw=ModifierStroke, inner sep=1.5pt, anchor=text, rectangle, rounded corners=1mm] {\color{ModifierText}{#1}};\phantom{#1\hspace{2pt}}}}}
}

\aptLtoXcmd{\long\def\circled#1{\xbox{aptbox}{\XMLaddatt{style}{background-color: \#000000; color: \#ffffff; border-radius: 50\%; width: 1.3em; height: 1.3em; display: inline-flex; align-items: center; justify-content: center; font-family: sans-serif; font-size: 0.75em; font-weight: bold; margin-left: -0.15em; margin-right: -0.1em; transform: translateY(-0.1em);font-style: italic;padding-right: 0.1em;padding-top: 0.1em;text-indent: 0;}#1}}}
{\DeclareRobustCommand*{\circled}[1]{\tikz[baseline=(number.base)]{\node[circle, fill=black, text=white, inner sep=0pt, minimum size=2.2ex, font=\sffamily\bfseries\itshape\footnotesize] (number) {#1};}}}

\newcommand{\rev}[1]{{#1}}

\makeatletter
\aptLtoXcmd{\renewcommand{\sectionautorefname}{\S}
}{
\renewcommand{\sectionautorefname}{\S\@gobble}
}
\makeatother

\begin{document}

\title{MuTable: Composable and Reusable Table Transformations for In-Situ Data Exploration}

\author{Fuling Sun}
\affiliation{
  \institution{University of California San Diego}
  \city{La Jolla}
  \state{California}
  \country{USA}
}
\email{fusun@ucsd.edu}

\author{Devamardeep Hayatpur}
\affiliation{
  \institution{University of California San Diego}
  \city{La Jolla}
  \state{California}
  \country{USA}
}
\email{dshayatpur@ucsd.edu}

\author{Jane L. E}
\affiliation{
  \institution{National University of Singapore}
  \city{Singapore}
  \country{Singapore}
}
\email{ejane@nus.edu.sg}

\author{Nicole Sultanum}
\affiliation{
  \institution{Tableau Research}
  \city{Seattle}
  \state{Washington}
  \country{USA}
}
\email{nsultanum@tableau.com}

\author{Haijun Xia}
\affiliation{
  \institution{University of California San Diego}
  \city{La Jolla}
  \state{California}
  \country{USA}
}
\email{haijunxia@ucsd.edu}


\begin{abstract}
Tables are central to data work to support precise lookup and full detail, but they can be limiting for overview and pattern-finding tasks. Visualizations are then created to gain richer perceptual support.
In practice, moving between tables and charts often requires maintaining parallel representations, introducing context switching, and extra coordination work. 
\rev{Building on prior hybrid table-visualization systems, we present \name, a prototype that reifies
transformations as persistent, composable, and reusable modifiers to support in-situ data exploration. Users can reshape the table while retaining and adapting intermediate forms as their questions evolve.}
An expert interview with eight data workers suggests that \name can support coordination between representations, rapid exploration, and greater user agency in constructing visualizations, \rev{as a low-commitment exploration space.}
\end{abstract}

\ccsdesc[500]{Human-centered computing~Visualization systems and tools}
\ccsdesc[500]{Human-centered computing~Interactive systems and tools}
\keywords{Direct Manipulation, Visualization, Table, Data Exploration}
\begin{teaserfigure}
  \includegraphics[width=\textwidth]{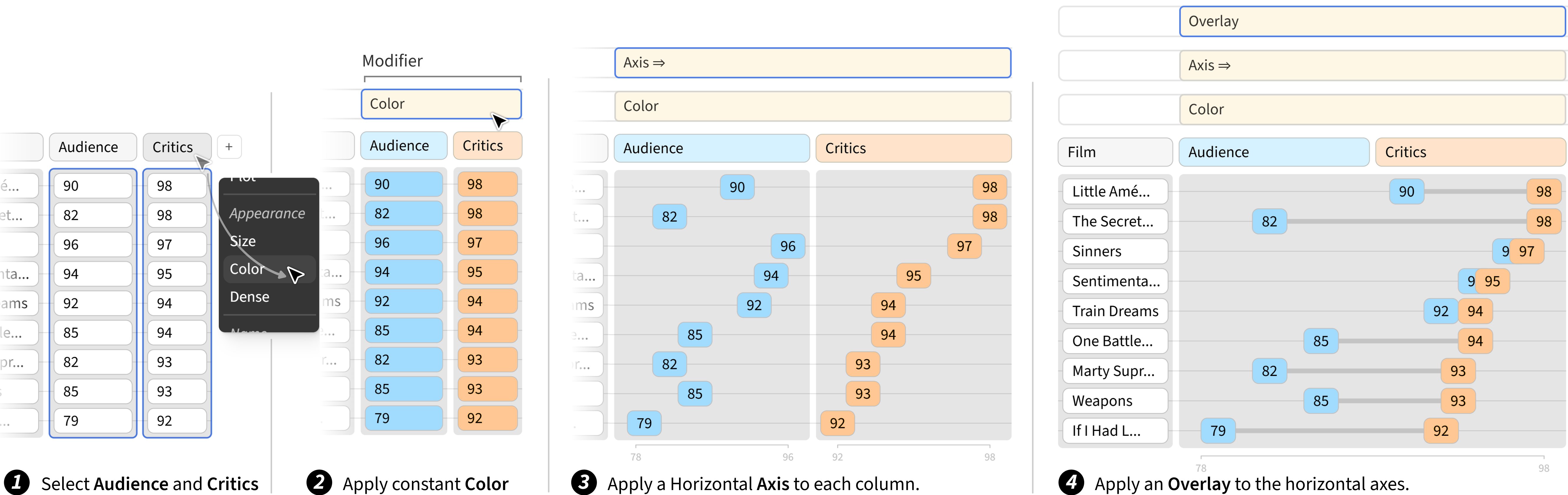}
  \caption{\name enables direct manipulation of tables into visualizations. Above, we partially replicate BBC's chart of 2018 Oscar nominations \cite{bbcOscars2018} for the 2026 Oscars. The user: \circled{1}) Selects the Audience and Critics columns; \circled{2}) Adds a \modifier{Color} \textit{modifier} on top of the two columns, modifiers are reified actions that span the columns they affect; \circled{3}) Adds a horizontal \modifier{Axis} to the columns, which places cells on a left-to-right axis based on their values; \circled{4}) Adds an \modifier{Overlay} to merge the two columns into a shared horizontal axis. (For brevity, only nine films with the highest critic ratings are shown.)}
  \Description{Four-panel example walkthrough showing how a table is transformed into visualization. In panel 1, the user selects the Audience and Critics columns in a table of film ratings. In panel 2, a constant color is applied, making Audience cells blue and Critics cells orange. In panel 3, a horizontal axis is applied to each column, repositioning the values as marks along separate numeric scales. In panel 4, the two axes are overlaid on a shared scale, with film titles at left and gray connectors linking each film’s Audience and Critics values.}
  \label{fig:teaser}
\end{teaserfigure}


\maketitle

\input{sections/01-Introduction}

\input{sections/02-Background}

\input{sections/03-Design}
\input{sections/04-User_Study}
\input{sections/05-Discussion}

\begin{acks}
We thank our anonymous reviewers, Matthew Beaudouin-Lafon, Emilia Rosselli Del Turco, Joshua Horowitz,  Jacob Yim, and the Foundation Interface Lab for their valuable input, and Bryan Min for suggesting the name \name. This work was supported by the National Science Foundation under Grant No. 2432644.
\end{acks}

\bibliographystyle{ACM-Reference-Format}
\bibliography{__references}

\clearpage
\onecolumn
\appendix
\input{sections/Appendix}
\end{document}

%% file: sections/01-Introduction.tex
\begin{figure*}[t]
    \centering
    \includegraphics[width=\linewidth]{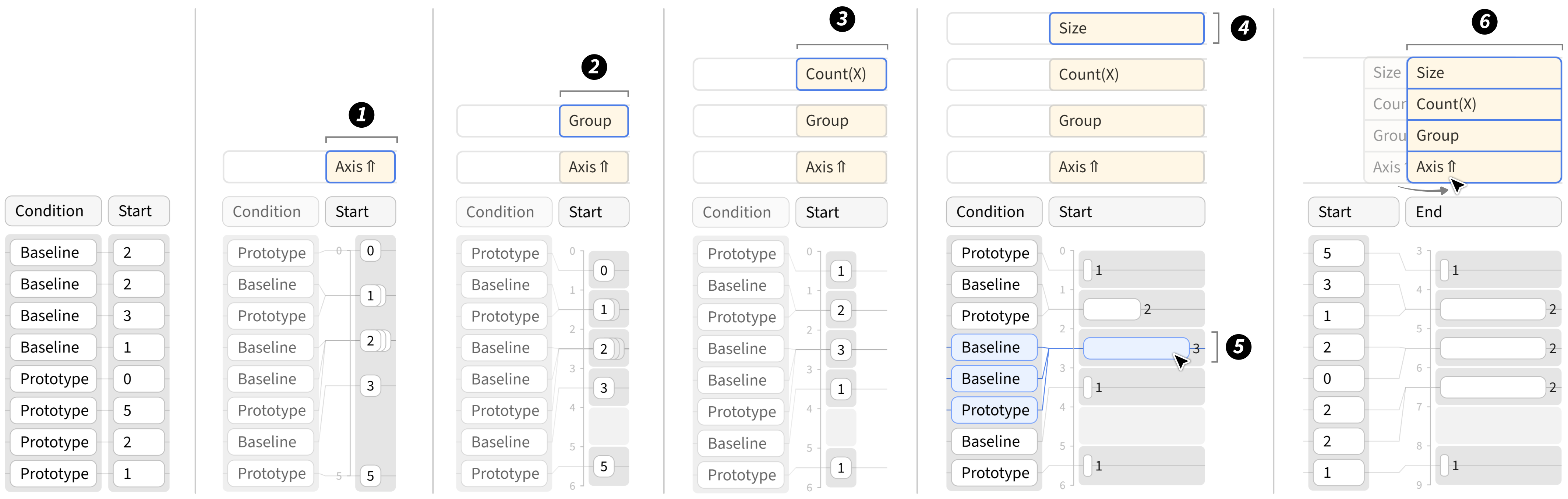}
    \caption{In \name, a histogram can be gradually specified in small steps. \circled{1}) Apply a vertical \modifier{Axis} to reposition cells on a vertical scale; \circled{2}) Apply \modifier{Group} to bin the cells along the scale; empty bins have a lighter background; \circled{3}) Count cells in each bin by applying \modifier{Count(X)}; \circled{4}) Apply \modifier{Size} to widen each cell by its value. \circled{5}) Hover over a bar to highlight the corresponding data in other columns. \circled{6}) The stacked modifiers can be joined into one unit, then reused by brushing it over other columns.}
    \Description{
    Six-panel walkthrough showing how a histogram is constructed from a table in MuTable. The source table has a categorical Condition column and a numeric Start column. In step 1, a vertical Axis is applied to the Start column, repositioning the numeric cells along a vertical scale. In step 2, a Group modifier bins the values along that axis, creating visible bins, with empty bins shown more faintly. In step 3, a Count(X) modifier aggregates the rows within each bin, replacing individual values with counts. In step 4, a Size modifier widens each aggregated cell according to its count, producing horizontal bars and forming a histogram. In step 5, hovering over one bar in the histogram highlights the corresponding rows in the Condition column. In step 6, the stack of modifiers is joined into a reusable unit and brushed onto another numeric column, End, to create another histogram using the existing modifiers.
    }
    \label{fig:histogram}
\end{figure*}

\section{Introduction}
\label{sec:intro}
Tables underpin data analyses from start to end and have served as a reliable format for recording, sharing, and presenting data for millennia~\cite{wright1970presenting, macdonald1977numbers}. 
They are familiar, and particularly well-suited for tasks that require direct data access and full context for precise value look-ups~\cite{munzner2014visualization}.
However, tables are ill-suited for overview-level and pattern-finding data tasks, areas where data visualizations and their use of pre-attentive visual attributes shine~\cite{saket2018task}.
\rev{
Therefore during data exploration, users often rely on both representations, moving between tables, for inspecting individual values, and visualizations, for identifying broader patterns to understand the data~\cite{idreos2015overview, tukey1977exploratory}.
}


\rev{
To reduce switching between views during exploration, prior work has sought to bring the tabular and visual structures together in \textit{hybrid} representations. 
For example, Table Lens uses visual encodings such as color and length to represent cell values~\cite{rao1994table}. Recent work further provides visual summaries (\eg, box plot, histogram, pie chart) of data attributes embedded as table cells~\cite{furmanova2020taggle, li2022hitailor}. Other systems compose individual tables and charts of data subsets with their connections visualized~\cite{viauConnectedChartsExplicitVisualization2012, gratzlDominoExtractingComparing2014}. These systems demonstrate how hybrid representations can reduce the separation between value inspection and higher-level visual analysis while preserving relationships among the represented data. }

\rev{
And yet, bringing tabular and visual forms together does not by itself support the iterative, improvisational nature of data exploration~\cite{alspaugh2018futzing}. 
Without knowing goals in advance, users may apply a transformation to inspect what it reveals, and then decide whether to extend it, revise it, or try another direction before finally committing to one. 
However, in many prior hybrid systems, transformations are transient commands, leaving the new representation visible while the operation that produced it is not. Users must reconstruct the previous steps to revisit alternatives and cannot easily adapt a transformation on another attribute.
}

\rev{
To address this gap, we present \name, a prototype system that reifies visual and data transformations as \emph{persistent, composable, reusable} modifiers to construct hybrid table-visualization representations during exploration. 
By keeping these operations \emph{persistent} alongside the evolving representation, users can revisit how the current view was produced and continue developing it (\autoref{fig:teaser}).
}
\rev{\name's interaction design follows two principles:}
\begin{enumerate}[leftmargin=*]
\item \textit{Small, composable operations.} 
Conventional spreadsheet tools provide a convenient, but non-customizable selection of preset charts. 
In contrast, \name supports flexible and fine-grained visual encoding manipulation centered around three categories: \textit{style}, \textit{layout}, and \textit{computation}, drawing on prior work~\cite{munzner2014visualization, wilkinson2011grammar}. 
\item \textit{Reified, reusable actions. } Conventional spreadsheets bake actions into the representation, \eg, sorting a table or creating a chart. It is difficult to revisit a previous operation, explore alternatives, or reuse the same transformation in other contexts. \name reifies actions as explicit, persistent objects called \textit{modifiers} that are layered on top of the table. Modifiers can be moved, extended, and stacked to support rapid composition. 
\end{enumerate}
\rev{Through persistent, composable, and reusable modifiers, \name provides a low-commitment space for exploring and becoming familiar with data before moving to focused analysis. Users can inspect, revise, and extend intermediate forms within the table as new questions emerge. 
This continuity positions \textit{data representations as materials} with which users can have \textit{reflective conversations}~\cite{schon1992designing}: users act on the table, interpret what each change reveals, and iteratively refine both their understanding of the data and the questions they pursue through persistent modifiers.
}
Our work contributes: 
\begin{itemize}[leftmargin=*]
    \item \rev{A table-centered interaction model, instantiated in \name, that reifies transformations as persistent, composable, and reusable modifiers, enabling users to construct hybrid table-visualization representations in-place incrementally to explore data. }
    \item An expert interview with eight data workers to investigate the utility of \name. The results show its potential for coordinating between tables and charts, rapidly exploring new data, and supporting user agency in constructing bespoke representations.
\end{itemize}

%% file: sections/02-Background.tex
\section{Background}
\label{sec:background}
We draw on several lines of work in HCI and visualization.

\paragraph{\rev{Hybrid Table-Visualization Systems}} 
\rev{Hybrid systems combine tabular and visual representations in several ways~\cite{furmanova2020taggle}. One approach treats tables and charts as connectable components, using visual links to show relationships across data subsets in separate views~\cite{gratzlDominoExtractingComparing2014, viauConnectedChartsExplicitVisualization2012}. 
\name builds on another approach, tabular visualization, which preserves the tabular layout while augmenting cells with visual encodings, such as bars and color fills~\cite{rao1994table, bertin1981graphics, perin2014bertifier}. Recent systems extend this approach with visual summaries of aggregated attributes (\eg, box plot, histogram,
pie chart)~\cite{furmanova2020taggle, li2022hitailor} or axis-based techniques (\eg, parallel coordinates) that reveal relationships across dimensions~\cite{gratzl2013lineup}. }

\rev{Users typically construct these forms through column-level configuration controls, such as selecting a visualization type from a panel associated with the target column~\cite{furmanova2020taggle}. Applying the same transformation elsewhere requires users to repeat the action across columns. Bertifier supports rapid application through \textit{crossets}: controls arranged beside rows and columns that users sweep across to toggle a transformation on adjacent cells~\cite{perin2014bertifier}. 
Like Bertifier, \name emphasizes rapid application, but further preserves each action as a persistent modifier that users can adjust and reuse.}

\paragraph{\rev{Visualization Formalisms}} 
Starting with \citet{wilkinson2011grammar}'s \textit{Grammar of Graphics}, various formal specifications have proposed primitive building blocks, like axes, marks, visual encodings, that can be composed to create complex charts \cite{satyanarayan2016vega, pollock2025gofish, petricek2021composabledatavis}. \name's operations draw on this line of work, \eg, the modifier \modifier{Overlay} is inspired by \citet{petricek2021composabledatavis}'s \textit{Compost}. 
However, in \name, actions occur \textit{imperatively} on a persistent representation of the table. The user edits the table \textit{in-place} through a series of modifiers, rather than composing it from scratch. 
In this sense, \name's semantics lie closer to \textit{procedural} specifications like Mascot \cite{liu2025manipulable} and Chisel \cite{hayatpur2025chisel} which use a sequence of steps to gradually generate visualizations. 

\rev{However, directly adopting one of these formalisms from the outset could constrain \name's design by shifting the focus toward generating valid chart specifications rather than preserving visualized columns within their surrounding table context. Therefore, we curated a set of column-level modifiers from these established vocabularies of these formalisms, including styling, spatial layout, and computation, to explore the design of \name.}


\paragraph{Instruments and Reification}
\textit{Instrumental interaction} is an interaction model which extends direct manipulation by distinguishing between \textit{objects of interest} and the \textit{instruments} used to act on them~\cite{beaudouin2000instrumental, hutchins1986direct}.
A central principle is \textit{reification}: making transient commands into explicit and persistent objects~\cite{beaudouin2000reification, beaudouin2000instrumental}. For example, a webpage scrollbar reifies the page navigation, while StickyLines turns the spatial alignment of graphical objects into visible and manipulable interface elements~\cite{ciolfi2016beyond}.
\name similarly reifies the actions on the table as \textit{modifiers}, enabling users to inspect, change, and compose actions together. We also support \textit{reuse}, as a modifier or composition of modifiers can be reapplied across columns without repeating actions from scratch.

%% file: sections/03-Design.tex
\begin{figure}[t]
    \centering
    \includegraphics[width=\linewidth]{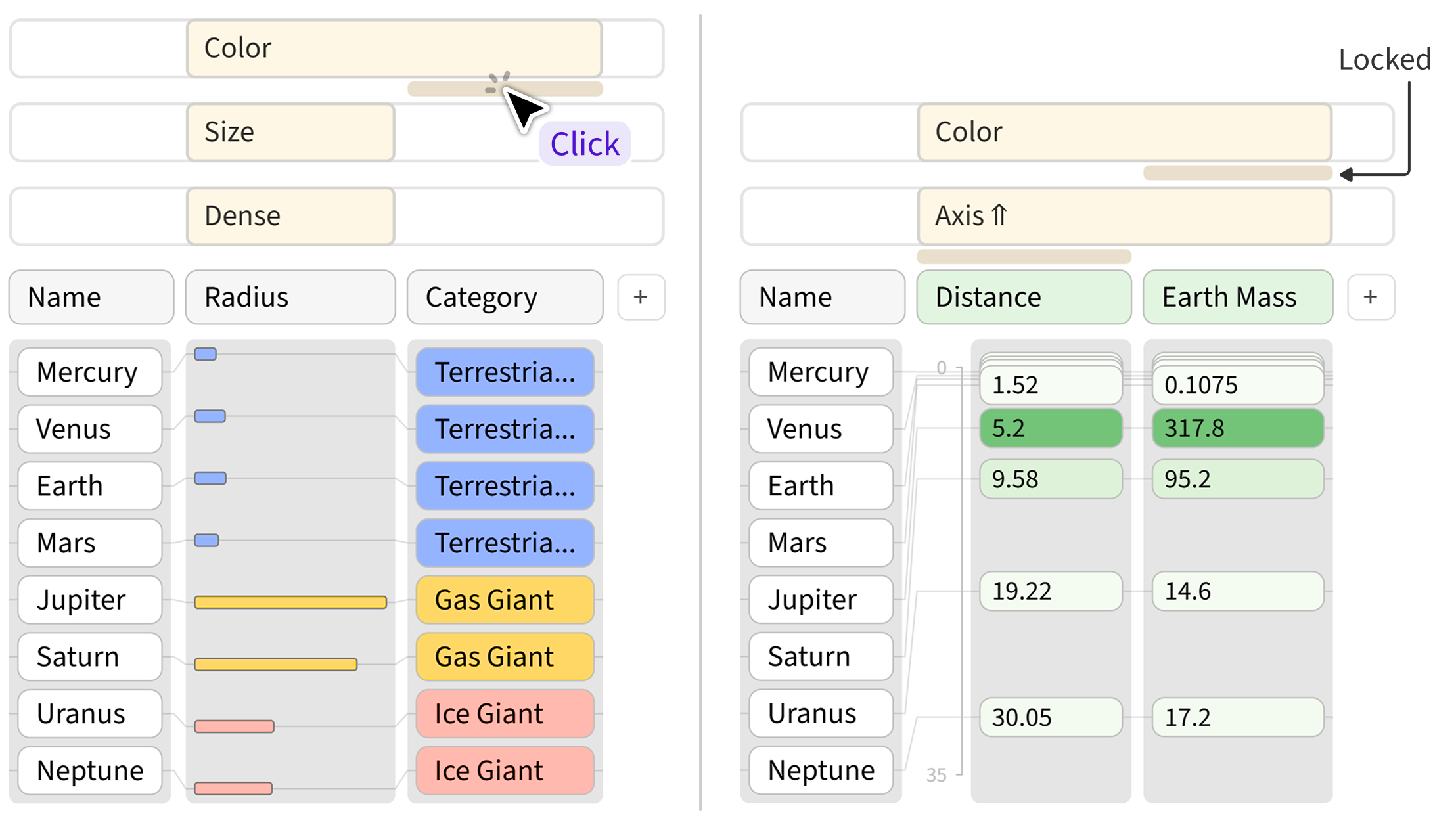}
    \caption{
    Modifiers can take one column as input and apply the resulting effect to others by ``locking''. Left: The Category column is locked in \modifier{Color}, applying its colors to Radius. Right: Distance is locked in a vertical \modifier{Axis} and Earth Mass in \modifier{Color}, then the columns share position and color.
    }
    \Description{Two examples showing how modifiers can lock one column as input and apply its effect to other columns. In the left panel, the Category column is locked into the Color modifier, causing the Radius column to inherit category-based colors: terrestrial planets are blue, gas giants are yellow, and ice giants are pink. In the right panel, the Distance column is locked into a vertical Axis, and the Earth Mass column is locked into a Color modifier. As a result, the Distance and Earth Mass columns share the same vertical positions and color encoding, allowing the two attributes to be compared through aligned placement and colors.}
    \label{fig:locking}
\end{figure}

\begin{figure*}[t]
    \centering   \includegraphics[width=\linewidth]{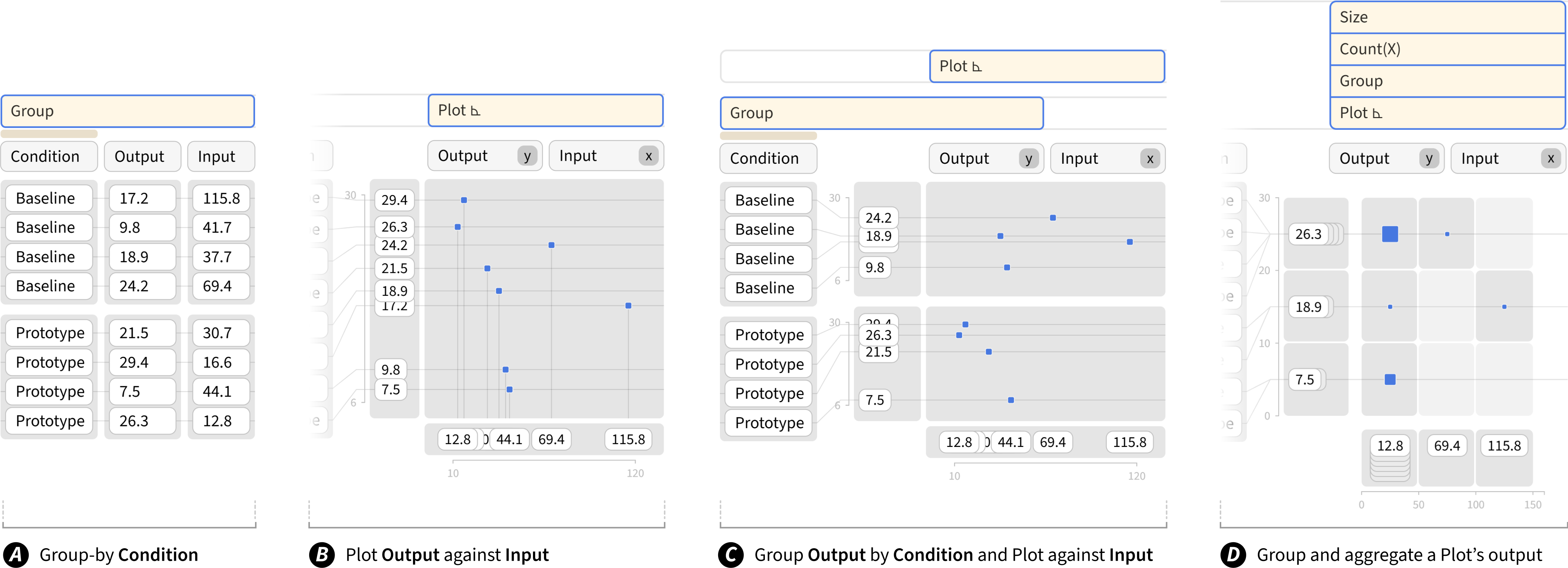}
    \caption{Different configurations of the same table. \circled{A}) \modifier{Group} with a locked categorical column partitions the other columns. \circled{B}) \modifier{Plot} two columns shows a scatterplot. \circled{C}) \modifier{Group} the y-column of the plot by the category results in stacked scatterplots for each condition. \circled{D})  \modifier{Group} and \modifier{Count(X)} a scatterplot bin data along both axes; \modifier{Size} then produces a \textit{bubble heatmap}.}
    \Description{Four configurations of the same table are shown. In panel A, grouping by the categorical column Condition partitions the rows into separate Baseline and Prototype sections while keeping the numeric Output and Input columns aligned within each group. In panel B, Output and Input are plotted as a scatterplot, with Output on the vertical axis and Input on the horizontal axis. In panel C, the scatterplot's y axis is grouped by Condition, producing stacked scatterplots for Baseline and Prototype that share the same horizontal scale. In panel D, the scatterplot is grouped and aggregated with Count(X) along both axes, and Size is applied to show the number of points in each bin, producing a bubble heatmap.}
    \label{fig:plot_and_group}
\end{figure*}

\section{\name}
Here, we describe the design of \name. \name is implemented as a web application, and can be accessed online.\footnote{\url{https://mutable-vis.github.io/}}

\subsection{Modifiers}
An action taken in \name is reified as a modifier. Modifiers are stacked above the table and operate on the columns they span. By default, a modifier acts on each column individually. For example, \autoref{fig:teaser} \circled{2} shows a \modifier{Color} modifier acting on the two columns beneath it. Once created, modifiers can be...
\begin{enumerate}[leftmargin=*]
    \item \textit{...reused} by moving or extending them to apply the same action to other columns. 
    \item \textit{...re-applied} by locking a column as an input so that other columns covered by the modifier will inherit the resulting effect (\autoref{fig:locking}). Locked columns move with the modifier.
    \item \textit{...joined} with other modifiers to form a single unit that can be reused as a sequence of actions (\autoref{fig:histogram} \circled{6}). 
\end{enumerate}

Modifiers can also be dragged vertically to reorder. Some actions behave differently based on their ordering, \eg, grouping \textit{then} aggregating is not the same as aggregating \textit{then} grouping.

\subsection{Actions}

\name supports various \textit{style}, \textit{layout}, and \textit{computation} actions. 
\rev{
These categories reflect progressively deeper interventions: \textit{style} modifiers change visual properties while preserving cell positions within the column; \textit{layout} modifiers reorganize cells spatially; \textit{computation} modifiers derive new values. 
Together they capture common visualization operations~\cite{wilkinson2011grammar, munzner2014visualization}, and many transformations in prior hybrid systems~\cite{rao1994table, furmanova2020taggle, gratzl2013lineup}. 
We selected representative modifiers from each category to explore the design of \name.
}

\subsubsection{Style}
Style actions change the appearance of cells. They include: \modifier{Color}, \modifier{Size} (\ie, cell width and/or height), and \modifier{Dense}. Cell color can be conditionally formatted based on values within a column by using a distinct hue for categorical data (\autoref{fig:locking}-left), and a gradient for numerical data (\autoref{fig:locking}-right). Or, it can be set to a constant color for all cells within the column (\autoref{fig:teaser}~\circled{2}). Cell size maps numerical values to size. \modifier{Color} and \modifier{Size} can reveal variation in data that are otherwise obscured in a generic table. When exact values are no longer needed, users can \modifier{Dense} cells, replacing the cell with a graphical mark. This reduces visual clutter and shifts attention toward broader visual patterns \cite{rao1994table, furmanova2020taggle}. \autoref{fig:locking}-left shows an example of composing \modifier{Dense}, \modifier{Size} and \modifier{Color}.

\subsubsection{Layout}
Layout actions include repositioning cells, grouping cells, and overlaying existing layouts. 

\paragraph{Reposition.} Cells can be repositioned by \modifier{Sort}ing them or by placing them on a vertical or horizontal \modifier{Axis}, which acts as a lightweight way to see the distribution of values within a column. When cells are vertically repositioned, the rest of the columns not being repositioned are reordered to keep the row-relations tidy (\eg, in \autoref{fig:histogram}~\circled{1}, after applying \modifier{Axis}, the Condition column is also reordered).
Often, it is useful to compare the distribution between two columns. A \modifier{Plot} modifier takes two columns as axes and generates a scatter plot. The cells in the two input columns are still visible and shown adjacent to the plot (\autoref{fig:plot_and_group}~\circled{B}).

\paragraph{Group.} A \modifier{Group} partitions a column by binning numerical values into ranges or separating unique categorical values. A group can be composed alongside other layouts to create small multiples (\autoref{fig:plot_and_group} \circled{C}), or used for later aggregation (\autoref{fig:histogram} \circled{3}).

\paragraph{Overlay.} An \modifier{Overlay} infers the range of cell positions set by an \modifier{Axis} and combines them into a shared axis \cite{petricek2021composabledatavis}, which is useful to compare columns of the same data domain (\autoref{fig:teaser}~\circled{3} \circled{4}).

\subsubsection{Computation}
Computation is useful for constructing visualizations (\eg, aggregating for creating a histogram in \autoref{fig:histogram}) and  \textit{descriptive summarization} (\eg, counting, summing, and averaging). To support computation, \name provides a small formula language with common mathematical operations (sum, count, natural logarithm, etc). Aggregations are performed at the group level, \eg, applying \modifier{Count(X)} atop a \modifier{Group} returns the number of items per-group (\autoref{fig:histogram}~\circled{3}, \autoref{fig:plot_and_group}~\circled{D}). Here, \texttt{X} refers to the column the computation is applied to. 
Formulas can also include absolute references to columns via their names. 
For example, in \autoref{fig:teaser}, a user could create a new column with a formula $\texttt{Critics - Audience}$ to compute and sort the films by the discrepancy in ratings.

\subsubsection{Composition} \name's small set of actions is versatile and composable to give access to a broad space of visualizations. For example, \autoref{fig:plot_and_group} shows various compositions of \modifier{Group} and \modifier{Plot}. In general, \name stores each column with a set of perceptual encodings (\eg, x, y, color, size), and a modifier can gain ownership over one or more of these encodings. For example, a horizontal \modifier{Axis} and a vertical \modifier{Axis} can compose to achieve a 2D layout similar to the \modifier{Plot} (given that the two axes have different locked columns).



\subsection{Using \name}
Below are three major usage patterns supported by \name. 
\paragraph{Brushing for rapid exploration.} 
\name's modifiers can be rapidly dragged and extended to other columns. 
For example, a user can apply a vertical \modifier{Axis} to one column, and then brush the modifier over other columns to quickly inspect each column's distribution, similar to the interaction in Magic Lenses~\cite{bier1993toolglassesandmagiclenses}.
Further, locking a column enables easy comparison across columns. For example, a user can lock the y input in \modifier{Plot} and drag the modifier to rapidly vary its x input to compare relationships among attributes.


\paragraph{Assembling and reassembling} \name's small, composable steps allow for gradual construction of visualizations. For example, \autoref{fig:histogram} demonstrates gradually building up a histogram. The individual steps (vertical axis, group, count, aggregate, and size) are both \textit{simple} and \textit{reappropriable}, they can be reassembled and modified to, \eg, create a bubble heatmap (\autoref{fig:plot_and_group}~\circled{D}).

\paragraph{Visible, manipulable, history} \name's actions in the form of modifiers can provide an accessible history for presenting and collaborating. For example, if a user were to share \autoref{fig:plot_and_group}~\circled{C} with a colleague, they would also be able to see the process to create the chart through the modifiers above the table. 

%% file: sections/04-User_Study.tex
\section{Expert Interview}
We conducted an expert interview to understand \name's perceived utility, limitations, and potential roles in data workflows. We recruited 8 data workers (P1--P8; 4 female; age 23--31) through department mailing lists and an online visualization community; all reported 5--10 years of data experience. All sessions were conducted over Zoom, lasted about 60 minutes, and compensated with a \$40 gift card.
Each session included a discussion of participants' data practices, a tutorial, a replication and open-ended exploration task, and a post-interview. Study details are provided in the \aptLtoX{Appendix \ref{app:interview}}{\autoref{app:interview}}.


\subsection{Findings}
From the thematic analysis (\aptLtoX{Appendix \ref{app:interview}}{\autoref{app:interview}}), we identified four themes.

\subsubsection{\name brings tables and charts closer.}
\label{sec:findingscloser}
When describing their prior experience, all participants noted shifting back-and-forth between tables and charts in their workflows (\eg, tracing outliers from a chart back to the table by P4 and P6). This segmented their workflow: \textit{``the exploratory analysis is very divorced from the data file [...] the visualization is not part of the file''} (P5).
To coordinate between tables and charts, participants relied on ad-hoc methods such as keyword searching (P6) or placing tables and charts side-by-side (P1, P3).
The distance between tables and charts can cause the two representations to drift: \textit{``a lot of the configuration that you do is from the chart. [But] from the table, you can't really see what operations each dimension has been used [for]''} (P8). 
Participants perceived \name as closing the distance between tables and charts (P1--P3, P5, P8). As P1 mentioned, \textit{``they are more interconnected, and I can directly visualize each column or different columns together.''}


\subsubsection{\name supports rapid, low-commitment exploration}
\label{sec:findingsrapid}
All participants noted \name's potential in exploring data (P1--P8), and P2, P5, and P6 emphasized the importance of freely exploring the data before committing to a particular analysis.
However, such exploration was perceived as costly in existing tools. Creating charts could require extra setup, such as duplicating data (P3) or preparing the data in a certain format (P6).
Even after a chart was created, modifying it could still be cumbersome (P1).
\name was perceived as lowering this cost by enabling quick, in-place transformations. P5 described dragging a modifier across columns to \textit{``quickly look what they all look like [in distribution] and then go back,''} while P6 noted that \textit{``it makes it easier for people to look at distributions, which is something I feel like scientists don't do enough.''}
Rapid and reversible exploration was also seen as valuable in collaboration. P3 and P4 described using \name in meetings, where visualizations could be adjusted rapidly in response to questions: \textit{``[if] your advisor asks, `what about this and this?' And you're just [...] dragging the modifiers [...] I do like these sorts of instantaneous visualizations for being able to answer a question on the fly''} (P4).



\subsubsection{\name supports user agency in constructing representations.}
\label{sec:findingsconstruct}
Participants noted the difference in granularity of actions when comparing \name to other spreadsheet tools.
For example, \textit{``one click for everything [in a spreadsheet] means you don't know what happened, what [are] the seven steps [that] happened behind this one click''} (P6). \name's incremental operations were helpful for understanding: \textit{``[I] feel confident when I'm looking at this chart because I created one in every single step''} (P8). Participants also valued the flexibility of composing these steps together to create visualization beyond templates (\eg, \textit{``to make funky combinations,''} P5), which was also beneficial for education purpose (P2, P6, P8). P6 contrasted \name with JASP (an open-source statistics program with analysis templates) that has predefined \textit{``format it [JASP] wants the data to be in.''}
%
\name was also perceived to give more direct control over the outcome. Instead of selecting from templates or \textit{``arguing with AI''} (P5), with \name, P5 noted that \textit{``it's just more like my hands are in the mud, like I'm working directly with this data set, like I'm directly moving stuff around and visualizing on the fly,''} emphasizing the feeling of directness and agency \cite{hutchins1986direct}.




\subsubsection{\name's limitations}
\label{sec:findingslimitation}
\rev{Participants raised concerns about scalability. For larger datasets, visual connections and layouts could introduce clutter (P6, P7); P6 suggested showing row connections only when needed. Modifier stacks could also consume substantial screen space (P4, P7); P4 proposed grouping them into collapsible compounds. We discuss these limitations further in \autoref{sec:limitations}.}

%% file: sections/05-Discussion.tex
\section{Discussion and Future Work}
We discuss \name's connections to other work and its limitations.

\subsection{Applications of \name}
We discuss three use cases identified in the study below.

\paragraph{Exploratory data analysis} \name's potential for rapid, low-commitment exploration (\autoref{sec:findingsrapid}) can be a boon to exploratory data analysis (EDA). EDA, broadly, is an attempt to answer ``what is going on here?'' when given novel data, with an emphasis on graphical representations and hypothesis generation~\cite{behrens1997principles}. 

\paragraph{Collaboration} \name may support collaborative work (\autoref{sec:findingsrapid}). Data work often involves communicating with stakeholders who vary in their familiarity with data analysis. To share reproducible analyses, data scientists often rely on programming tools such as Jupyter Notebooks, though these often require careful curation for general audiences~\cite{wang2019data,rule2018exploration}. \name could offer a more accessible alternative for non-programmers.



\paragraph{Learning by Constructing}
\name promotes agency: users construct visualizations step-by-step, maintaining control throughout (\autoref{sec:findingsconstruct}). 
This style of interaction aligns closely with Constructive Visualization~\cite{huron2014constructive}, a visualization authoring paradigm in which users assemble visualizations from basic units (\eg, tangible blocks~\cite{fan2020constructive}) by gradually changing their visual properties (\eg, color, size) and arrangement. Constructive Visualization emphasizes \textit{simplicity} (\ie, a small set of steps), \textit{expressivity} (\ie, sufficient freedom to ``assemble signs''), and \textit{dynamism} (\ie, can be ``easily rebuilt and adjusted'') -- all three of which are emphasized in the design of \name. 



\subsection{Data Representations as Materials}
\name places tables and charts in a shared space, reducing context switches and supporting coordination across representations (\autoref{sec:findingscloser}).
\rev{This ability to reshape a representation in place suggests viewing \textit{data representations as materials} for reflective conversation~\cite{schon1992designing}, where each transformation not only reshapes the representation but also reveals consequences that can guide the user's next action. Prior work has explored such continuity in:}
\begin{enumerate}[leftmargin=*]
    \item \textit{...transitions} between representations to help users form analogies~\cite{ruchikachorn2015learning} and understand operations like aggregation \cite{kim2019designing},
    \item \textit{...folding and unfolding} information by fluidly changing the level of detail, to support visual data exploration \cite{bludau2025fluidly},
    \item \textit{...semantic zooming} to dynamically scale data items  \cite{bederson1996pad++, rao1994table, cockburn2009review},
    \item \textit{...sculpting} to programmatically manipulate data in-place \cite{horowitz2025sculpin}.
\end{enumerate}

\rev{This perspective can be extended to other representations and practices. }
For example, qualitative analysis often spans canvases, text, and tables, which could be integrated into an evolving medium.





\subsection{Limitations}
\label{sec:limitations}

\rev{
First, \name faces scalability challenges along two dimensions. At the data level, larger datasets introduce perceptual and computational challenges~\cite{Wickham2013BinsummarisesmoothA, feketeHumanDataInteractionExploration2026}. 
Future work can combine data reduction and abstraction techniques such as filtering, sampling, and aggregation, with scalable visualization techniques~\cite{mayorga2013splatterplots, furnas1986generalized}.
Since \name preserves both tabular and visual representations, these strategies can be designed as modifiers and applied during exploration. 
At the interaction level, tall stacks of modifiers may consume substantial screen space and become difficult to navigate. Potential approaches include collapsing related modifiers, representing inactive operations as compact summaries, and supporting search and branching. Future work should examine how these mechanisms can reduce complexity without obscuring the provenance of transformations.}

\rev{Second, the current prototype supports a limited set of operations, reflecting our focus on the interaction model rather than a comprehensive visualization grammar. Future work could draw on more expressive formalisms, such as Mascot~\cite{liu2025manipulable}, to expand the range of marks, layouts, and data transformations available as modifiers.
The modifier vocabulary could also be extensible, allowing users to define and incorporate domain-specific operations.} 

Finally, we did not evaluate \name in the context of a practical workflow, or over extended periods. A longitudinal study will provide more concrete insights into how users adapt \name to their existing workflows, \rev{which modifier sequences they repeatedly reuse, and where friction arises when moving from \name to more specialized downstream tools. }

\section{Conclusion}

Tables are central to data work, yet analysts often leave them to gain richer perceptual support from visualizations. We present \name, \rev{a prototype that supports the progressive construction of hybrid table-visualization representations during exploratory data work using persistent, composable, and reusable modifiers.}
Our expert interview suggests that this approach better coordinates tables and charts, supports fluid exploration, and gives users greater agency in visualization construction. 
\rev{More broadly, this work views data representations as materials that users can progressively reshape in place, preserving continuity as their understanding develops.}

%% file: sections/Appendix.tex
\section{Expert Interview}
\label{app:interview}
\subsection{Study Procedure}
\rev{We designed the study to examine how participants understand and engage with \name, rather than to directly compare task performance against existing tools. We focused on gathering qualitative feedback on the system's perceived utility, limitations, and potential roles within participants' existing workflows. 
Each session began with informed consent, after which the study proceeded as follows:}
\begin{enumerate}[leftmargin=*]
    \item \textit{Introduction (10 min)}. Participants were asked to describe their experience working with data, share a recent data-related task they had performed, and briefly reflect on how they currently coordinate between tables and charts.
    \item \textit{\name Tutorial (20 min)}. Participants were provided a tutorial through \name's features \rev{using sample datasets and scenarios}. We introduced modifiers progressively, beginning with more familiar ones (\eg, color and size) and then moving to more novel ones (\eg, overlay) and their composition. \rev{Participants were encouraged to ask questions and share their comments on \name during the process.}
    \item \textit{Replication and Open-ended Exploration (15 min)}. Participants were guided through recreating an existing visualization~\cite{bbcOscars2018} (similar to \autoref{fig:teaser}), with assistance from the interviewer when needed. After completing the replication task, participants conducted open-ended exploration on sample datasets. \rev{This stage was designed to examine whether participants could understand and use modifiers compositionally, while providing a concrete task and approachable dataset through which to engage with \name and offer feedback.}
    \item \textit{Interview (15 min)}. A post-interview probed participants' impressions of \name's design, including embedding different visual representations within tables and the use of modifiers. \rev{Participants were encouraged to continue interacting with \name during the interview so they could ground their feedback in specific examples.} 
\end{enumerate}
\subsection{Participant Information}
Participants were 23--31 years old ($M=27.75, SD=2.60$), and reported 5--10 years of experience with data analysis ($M=7.5, SD=1.93$). \autoref{tab:participants} provides detailed participant information.
\begin{table*}[!h]
\caption{Participant information. ID indicates participant ID, and Years indicates years of experience with data analysis.}
\centering
\small
\setlength{\tabcolsep}{4pt}
\renewcommand{\arraystretch}{1.15}
\begin{tabularx}{\textwidth}{l l l l l l >{\raggedright\arraybackslash}X}
\toprule
\textbf{ID} & \textbf{Age} & \textbf{Gender} & \textbf{Occupation} & \textbf{Domain} & \textbf{Years } & \textbf{Data Analysis Tools Used} \\
\midrule
P1 & 23 & Female & Ph.D. Student & HCI & 6 & Excel, Google Sheets, Tableau / Power BI, Python (pandas, etc.) \\
P2 & 31 & Female & Designer & UX/UI Design & 5 & Excel, Google Sheets \\
P3 & 29 & Female & Ph.D. Student & Cognitive Science & 7 & Excel, Google Sheets, Python (pandas, etc.) \\
P4 & 30 & Male & Ph.D. Student & Neuroscience & 10 & Google Sheets, Python (pandas, etc.) \\
P5 & 28 & Male & Ph.D. Student & Neuroscience & 9 & Excel, Google Sheets, Python (pandas, etc.) \\
P6 & 26 & Female & Ph.D. Student & Cognitive Science & 6 & Google Sheets, Python (pandas, etc.) \\
P7 & 26 & Male & Ph.D. Student & Neuroscience & 7 & Excel, Google Sheets, Python (pandas, etc.), R \\
P8 & 29 & Male & Ph.D. Student & Visualization & 10 & Excel, Google Sheets, Tableau / Power BI, Python (pandas, etc.), R \\
\bottomrule

\end{tabularx}

\label{tab:participants}
\end{table*}

\subsection{Analysis}
The first author conducted a thematic analysis of the interview data by reviewing the interview recordings alongside notes taken during each session. Themes were iteratively developed, grouped, and refined to capture participants' perceptions of \name's utility, limitations, and potential roles within existing data workflows.